\documentclass[letterpaper]{jpconf}
\newcommand{\be}{\begin{equation}}
\newcommand{\ee}{\end{equation}}
\newcommand{\ba}{\begin{eqnarray}}
\newcommand{\ea}{\end{eqnarray}}
\begin{document}
\title{Lorentz violating electrodynamics}

\author{Belinka Gonz\'alez$^1$, Santiago A Mart\'\i nez$^2$, Rafael Montemayor$^2$
and Luis F Urrutia$^1$}

\address{$^1$ Instituto de Ciencias Nucleares,
Universidad Nacional Aut{\'o}noma de M{\'e}xico,  A. Postal
70-543, 04510 M{\'e}xico D.F., M{\'e}xico}
\address{$^2$ Instituto Balseiro and CAB, Universidad Nacional de Cuyo and CNEA,
8400 Bariloche, Argentina}

\ead{$^1$urrutia@nucleares.unam.mx, $^2$montemay@cab.cnea.gov.ar}

\begin{abstract}
After summarizing the most interesting results in the calculation
of synchrotron radiation in the Myers-Pospelov  effective model
for Lorentz invariance  violating (LIV) electrodynamics, we
present a general unified way of describing the radiation regime
of LIV electrodynamics which include the following three different
models : Gambini-Pullin, Ellis et al. and Myers-Pospelov. Such
unification reduces to the standard approach of radiation in a
dispersive and absortive (in general) medium  with a given index
of refraction. The formulation is presented up to second order in
the LIV parameter and it is explicitly applied to the synchrotron
radiation case.
\end{abstract}

\section{Introduction}
The appearance of detectable low energy effects arising from
possible quantum gravity corrections to particle dynamics
\cite{ACNAT} has been recently the subject of intense scrutiny
\cite{analysis,JACOBSON}. The most direct interpretation of such
effects, though not the only one \cite{ADDINT}, is related to the
breaking of observer Lorentz covariance. In this way this subject
becomes directly tied to the long time honored investigations,
both theoretical and experimental, based upon the Standard Model
Extension \cite{KOSTELECKY} concerning Lorentz and CPT violations
\cite{CPTMEETS}. Heuristic derivations of such effects from loop
quantum gravity have been presented \cite{GPED,URRU}, which make
clear that a better understanding of the corresponding
semiclassical limit is required \cite{THIEMANN}. String theory has
also provided models for explaining such quantum gravity induced
corrections \cite{QFMODEL}. Also, effective field theory models to
describe such effects have been constructed \cite{MP}.

The observation of $100\, MeV$ synchrotron radiation from the Crab
Nebula has recently been used to impose very stringent limits upon
the parameters describing a modified photon dynamics, embodied in
Maxwell equations that get correction terms which are linear in
the Planck length \cite{JACOBSON}. Such bounds are based on a set
of very reasonable assumptions on how some of the standard results
of synchrotron radiation extend to the Lorentz non-invariant
situation. This certainly implies some dynamical assumptions,
which have been recently tested  by two of us (RM and LFU)
 \cite{MU} in the framework  of the  classical version of the
Myers-Pospelov (MP) effective theory, which parameterizes LIV
using dimension five operators, as described in the first paper of
Ref. \cite{MP}. Ref. \cite{MU} summarizes a complete calculation
of synchrotron radiation in the context of this model. This
constitutes an interesting problem on its own whose resolution
will subsequently allow the use of additional observational
information to put bounds upon the correction parameters. For
example we have in mind the polarization measurements from
cosmological sources. The case of gamma ray bursts  has recently
become increasingly relevant \cite{CB}, although it is still at a
controversial stage \cite{CONTR}.

This paper is organized as follows. In Section 2, we provide some
highlights of the results obtained in Ref. \cite{MU}, and from
there onwards we develop a formalism for a Lorentz violating
electrodynamics based on extended constitutive relations for the
$\mathbf{D}$ and $\mathbf{H}$. In Section 3 we state the general
structure of this theory, in Section 4 we introduce a
parameterization for the constitutive relations, and in Sections 5
and 6 we summarize the main expressions for the fields, energy
spectrum, power spectrum and polarization in terms of effective
refraction indices. We close with a short summary in section 7.

\section{Synchrotron radiation in the Myers-Pospelov model}

This model parameterizes LIV using dimension five operators both
in the matter and electromagnetic sectors. There is also a
preferred frame four velocity $V^\mu$, which is not a dynamical
field. We choose to work in the rest frame $V^\mu=(1,\mathbf{0})$.
As usual the model possesses observer Lorentz covariance, which
means that the fields and the four-velocity $V^\mu$ transform as
tensors  when going from one observer frame to another. On the
other hand, in each frame we violate particle Lorentz
transformations; that is to say  we have a collection of
non-dynamical physical fields in the action,  in analogy to the
physics going on in the presence of an external magnetic field
which violates particle rotation invariance, for example.

In the rest frame, the modified Maxwell equations are ($c=1$)
\ba
&&\nabla \cdot \mathbf{B}=0,  \qquad\quad  \mathbf{\nabla \times
E+}\frac{\partial \mathbf{B}}{\partial t}=0,  \nonumber
\\
&&\nabla \cdot \mathbf{E}=4\pi\rho, \qquad  -\frac{\partial
\mathbf{E}}{\partial t}+\nabla \times \mathbf{B}+\xi
\partial _{0}\left( -\nabla \times \mathbf{E}+\partial _{0}\mathbf{B}%
)\right)=4\pi \mathbf{j},  \label{MP}
\ea%
where the LIV parameter $\tilde \xi$ is usually written as $\tilde
\xi= \xi/M$ with $\xi$ being a dimensionless number and $M$ being
a mass scale characterizing the Lorentz symmetry breaking, which
is usually, but not necessarily, identified with the Plank mass.
For the particular case of synchrotron radiation the modified
dynamics for a  particle of charge $q$ in an external constant
magnetic field in the $z$-direction produces the standard
helicoidal orbits with a modified Larmor frequency $\omega_0$
\begin{equation} \mathbf{\ddot{r}}=\frac{q}{E}\left(
1-\frac{3}{2}{\tilde \eta} E+\frac{9}{4} {\tilde \eta}^{2}
E^{2}\right)  \left( \mathbf{v\times B}\right)\equiv \omega_0
\left( \mathbf{v\times B}\right), \label{EQCHARGE}
\end{equation}
depending upon another LIV parameter $\tilde \eta= - \eta/M$ and
with $E$ being the energy of the particle. The minus sign is
chosen to make contact with the notation of Ref.\cite{JACOBSON}.
We restrict ourselves to circular orbits in the $x-y$ plane with
Larmor radius  $R=\beta/\omega_{0}$ and we use the standard
notation $\beta=|\mathbf{v|}$ and $\gamma=\left( 1-\beta^{2}
\right) ^{-1/2}$. When rewriting $\beta$ in terms of the particle
energy we obtain the modified expression
\begin{equation}
1-\beta^{2}(E)=\frac{\mu ^{2}}{E^{2}}\left[
1+2\frac{\tilde{\eta}E^{3}}{\mu
^{2}}-\frac{15}{4}\frac{\tilde{\eta}^{2}E^{4}}{\mu ^{2}}+O(\tilde{\eta}^{3})%
\right].  \label{UMB2}
\end{equation}%
Due to de admixture of vector and axial vectors in the last Eq.
(\ref{MP}) the problem presents birefringence and the two
independent modes of propagation correspond to those of definite
circular polarization $\lambda=\pm $, with refraction indices
$n_\lambda(z)$
\begin{equation}
n_\lambda(z)=\sqrt{1+z^{2}}+\lambda z, \quad  z=\tilde{\xi}\omega. \label{REFIND0}%
\end{equation}
Such modes are identified via the corresponding Green functions,
which are most easily calculated in the radiation gauge. It is
interesting to observe that Eqs. (\ref{MP}) provide the exact
expression \be
4\pi\mathbf{S}=\mathbf{E}\times\mathbf{B}-\tilde{\xi
}\mathbf{E}\times\partial\mathbf{E}/\partial t, \label{pv}%
\ee
for the Poynting vector. Also, the magnetic and electric field
satisfy the relation
\begin{equation}
\mathbf{B}(\omega,\mathbf{{\hat{n}}})=\sqrt{1+{z^{2}}}\,\mathbf{\hat{n}\times
E}(\omega,\mathbf{\hat{n}})-i{z}\,\mathbf{E}(\omega,\mathbf{\hat{n}}).
\label{BAFE}%
\end{equation}
in the radiation approximation, with
$\mathbf{\hat{n}}=\mathbf{r}/r$ being the direction of
observation. These two effects conspire to produce a positive
definite expression for the power flux yielding the following
expression for the average angular distribution of the radiated
power spectrum \ba &&\left\langle \frac{d^{2}P(T)}{d\omega
d\Omega}\right\rangle =\sum
_{\lambda=+,-}\,\sum_{m=0}^{\infty}\delta({\omega}-\omega_{m})\frac
{dP_{m,\,\lambda}}{d\Omega}, \quad \omega_{m}=m\omega_{0},
\quad z_{m}=\tilde{\xi}\omega_{m}, \label{HARMEXP} \\
&&\frac{dP_{m,\,\lambda}}{d\Omega}=\frac{\omega_{m}^{2}q^{2}}{4\pi}\frac
{1}{\sqrt{1+z_{m}^{2}}}\left[  \lambda \beta n_\lambda(
z_{m})J_{m}^{\,\prime }(W_{\lambda
m})+\,\cot\theta\,J_{m}(W_{\lambda m})\right]^{2}.
\label{ADMH}%
\ea Here, the average $\left\langle \right\rangle $ is taken with
respect to the macroscopic time $T$ and $J_{m}$,
$J_{m}^{\,\prime}$ denote the Bessel functions and their
derivatives respectively. The argument of the Bessel functions is
$ W_{\lambda m}=m\,n_\lambda( z_{m})\beta\sin\theta$. We also have
calculated the total averaged and integrated
power radiated into the $m$-th harmonic%
\be P_{m} =\frac{q^{2}\beta^{2}\omega_{m}}{R
\,\sqrt{1+z_{m}^{2}}}\,\sum_{\lambda=\pm
}n_\lambda( z_{m})\left[J_{2m}^{\,\prime}(2m\;n_\lambda( z_{m}%
)\beta) -\frac{1-\beta^{2}n_\lambda^{2}(
z_{m})}{2\beta^{2}n_\lambda^{2}( z_{m})}  \int
_{0}^{2mn_\lambda( z_{m})\beta}dx\;J_{2m}(x)\right]  , \label{PMDO}%
\ee
which clearly indicates the contribution of each polarization
$P_{\lambda m}$. The above result is exact in $z_{m}$ and the
parity-violating contribution has vanished after the angular
integration.

In the case of synchrotron radiation from Crab nebula, as well as
from other objects like Mkn 501 and GRB021206, one can estimate
that the $m$ corresponding to the observed radiation are very
large, let us say in the range $10^{15}-10^{30}$, with the
respective values of $m/\gamma$ varying within $10^{10}-10^{20}$.
These properties motivate the large $m$ and $m/\gamma$ expansion
of the Bessel functions involved in the spectrum. In a similar way
to the standard case \cite{SCWANNP}, a first consequence of this
approximation is the appearance  of the cutoff frequency
$\omega_{\lambda c}=\tilde{m}_{\lambda c}\,\omega_0$ with
\begin{equation}
\tilde{m}_{\lambda c}=\frac{3}{2}\left[  1-\beta^{2}(E)\,n_\lambda^{2}({z_{m}%
})\right]  ^{-3/2}. \label{HCOFF}%
\end{equation}
This means that for $m > \tilde{m}_{\lambda c}$ the total power
decreases as $P_{\lambda m} \approx \exp(-m/{\tilde{m}_{\lambda
c}})$. Within the same large-$m$ approximation, the integrated
power in the $m$-th harmonic can be expanded to second order in
${\tilde{\xi}}$ yielding%
\begin{equation}
P_{m}=\frac{q^{2}\omega}{\sqrt{3}\pi R \gamma^{2}}\left[  \frac{m_{c}}%
{m}{\kappa}\left(  \frac{m}{m_{c}}\right)  -\frac{2}{\gamma^{2}}%
K_{2/3}\left(  \frac{m}{m_{c}}\right)  +2\left(  \frac{{\tilde{\xi}}%
\,m\omega\beta}{\gamma}\right)  ^{2}K_{2/3}\left(
\frac{m}{m_{c}}\right)
\right]  , \label{PMEXP}%
\end{equation}
where $m_{c}=3\gamma^{3}/2$ and
${\kappa}(x)=x\int_x^{\infty}dy\,K_{5/3}(y)$ is the so called
bremsstrahlung function. Here $K_{p/q}$ denote the Bessel
functions of fractional order. Let us notice  the appearance of
the combination ${\tilde{\xi }}\,\omega
m/\gamma=\xi(\omega/M_{P})(m/\gamma)$ as the expansion parameter
in (\ref{PMEXP}). This is not necessarily a small number, which
signals the possibility that such corrections might be relevant in
setting bounds upon $\tilde{\xi}$. This rather unexpected effect
is due to the amplifying factor $ m/\gamma$. Similar results have
been obtained in calculations of the synchrotron radiation spectra
in the context of non-commutative electrodynamics\cite{Castorina}.

Another possibility for observable effects due to ${\tilde{\xi}}$
is to look
at the averaged degree of circular polarization%
\begin{equation}
\Pi_{\odot}=\frac{\left\langle
P_{+}(\omega)-P_{-}(\omega)\right\rangle
}{\left\langle P_{+}(\omega)+P_{-}(\omega)\right\rangle }, \label{CIRCPOL}%
\end{equation}
where $P_{\lambda}(\omega)$ is the total power distribution per
unit frequency
and polarization $\lambda$, so that $P_{\lambda}(\omega)=P_{m\lambda}%
/\omega_{0}$. The average here is calculated with respect to an
energy distribution of the relativistic electrons, which we take
to be $N(E)dE=CE^{-p}dE$, in some energy range $E_{1}<E<E_{2}$,
chosen  as $E_1=0$ and $E_2 \rightarrow \infty$
for simplicity. The result is%
\be
\Pi_{\odot}={\tilde{\xi}}\omega\left(
\frac{\mu\omega}{qB}\right) \,\Pi(p),
\label{CPEXP}%
\ee
\begin{equation}
\Pi(p)=\frac{(p-3)\left(  3p-1\right)  }{3\left(  3p-7\right)
}\,\frac {(p+1)}{(p-1)}\frac{\Gamma\left(
\frac{p}{4}+\frac{13}{12}\right)
\Gamma\left(  \frac{p}{4}+\frac{5}{12}\right)  }{\Gamma\left(  \frac{p}%
{4}+\frac{19}{12}\right)  \Gamma\left(
\frac{p}{4}+\frac{11}{12}\right)
},\quad p>7/3. \label{PIP}%
\end{equation}
Again, we have the presence of an amplifying factor in Eq.
(\ref{CPEXP}), given by $(\mu\omega/qB)$, which is independent of
the form of $\Pi(p)$ and not necessarily a small number.  An
estimation of this factor in the zeroth-order approximation
(${\tilde \xi}=0={\tilde \eta}$), which is appropriate in
 (\ref{CPEXP}), yields $(\mu\omega/q B )=\omega/(\omega_0
\gamma)=m/\gamma$. The expression (\ref{CPEXP}) is analogous to
the well-known average of the degree of linear polarization
$\Pi_{LIN}=(p+1)/(p+7/3)$, under the same energy distribution for
the electrons.

\section{General structure of Lorentz violating electrodynamics}

Three paradigmatic examples of Lorentz violating electrodynamics
are given by the effective theories proposed by Gambini and Pullin
(GP) \cite{GPED}, Ellis et al.(EMN)\cite{QFMODEL}, and Myers and
Pospelov (MP) \cite{MP}. They can be written in the general form
of Maxwell equations
\begin{eqnarray}
&&\mathbf{\nabla }\cdot \mathbf{D} =4\pi \rho
,\;\;\;\;\;\;\mathbf{\nabla }\cdot
\mathbf{B}=0,  \label{GENMAXW1}\\
&&\mathbf{\nabla }\times \mathbf{E} =-\frac{\partial \mathbf{B}}{\partial t}%
,\;\;\mathbf{\nabla }\times \mathbf{H}=\;\frac{\partial
\mathbf{D}}{\partial t}\;+4\pi \mathbf{j},  \label{GENMAXW2}
\end{eqnarray}%
with corresponding constitutive relations%
\begin{equation}
\mathbf{D=D(E,B),\;\;\;H=H(E,B)},  \label{GENCR}
\end{equation}%
which we next write in detail for each case, after reviewing the
corresponding equations. Let us recall that the above equations
(\ref{GENMAXW1}) and  (\ref{GENMAXW2}) imply charge conservation $
\partial \rho /\partial t\;+\mathbf{\nabla }\cdot
\mathbf{j}=0, $ independently of the constitutive equations
(\ref{GENCR}). In an abuse of notation have  denoted by ${\tilde
\xi}$ the electromagnetic LIV parameter for all models in the
sequel.

\subsection{Gambini-Pullin Electrodynamics}

The Maxwell equations for this case are%
\ba &&\mathbf{\nabla \cdot }\mathcal{B}=0, \qquad \;\;\;
\mathbf{\nabla }\times \left( \mathcal{E}+2{\tilde \xi}
\mathbf{\nabla }\times \mathcal{E}\right)
+\frac{\partial \mathcal{B}}{\partial t}=0 ,  \\
 &&\mathbf{\nabla \cdot }\mathcal{E}=4 \pi \rho, \qquad \mathbf{\nabla }\times \left( \mathcal{B}+2{\tilde \xi} \mathbf{\nabla
}\times \mathcal{B}\right) -\frac{\partial \mathcal{E}}{\partial
t}=4\pi \mathbf{j},   \label{GP}
\ea%
where the electric and magnetic fields are identified from the
homogeneous equation
as%
\begin{equation}
\mathbf{E}=\mathcal{E}+2{\tilde \xi} \mathbf{\nabla }\times
\mathcal{E},\qquad\mathbf{B}=\mathcal{B}.
\end{equation}%
From the inhomogeneous equations we obtain
\begin{equation}
\mathbf{D}=\mathcal{E}, \qquad \mathbf{H}=\mathcal{B}+2{\tilde
\xi} \mathbf{\nabla }\times \mathcal{B},
\end{equation}%
which  together with the constitutive relations
\begin{equation}
\mathbf{D}+2{\tilde \xi} \mathbf{\nabla }\times \mathbf{D=E}, \qquad \mathbf{H=B}+2{\tilde \xi} \mathbf{%
\nabla \times B},  \label{CRGP}
\end{equation}%
leave the equations in the required form. These equations define
the corresponding functions stated in
 (\ref{GENCR}). In momentum space%
\begin{equation}
\mathbf{D}=\frac{1}{1+4{\tilde \xi} ^{2}k^{2}}\left(
\mathbf{E}-2i{\tilde \xi} \mathbf{k }\times \mathbf{E}+4{\tilde
\xi} ^{2} \left( \mathbf{k } \cdot \mathbf{E}\right)\mathbf{k }
\right),\qquad \mathbf{H}=\mathbf{B}+2i{\tilde
\xi}\mathbf{k}\times\mathbf{B}.\label{GPCE}
\end{equation}%
The admixture of vectors and axial vectors in the constitutive
relations precludes the parity violation exhibited by  the model,
together with the presence of birefringence.

\subsection{Ellis et. al. Electrodynamics}

In this case the modified Maxwell equations are
\ba
&&\nabla \cdot \mathbf{B}=0, \qquad\qquad\qquad \mathbf{\nabla }\times
\mathbf{E}+\frac{\partial \mathbf{B}}{\partial t}=0,  \\
&&\mathbf{\nabla }\cdot \mathbf{E}+\mathbf{u}\cdot \frac{\partial \mathbf{E}}{%
\partial t}=4\pi \rho _{eff}=4\pi (\rho -\mathbf{u\cdot j}),   \\
&&\mathbf{\nabla }\times \mathbf{B}-\left( 1-\mathbf{u}^{2}\right) \frac{%
\partial \mathbf{E}}{\partial t}+\mathbf{u}\times \frac{\partial \mathbf{B}}{%
\partial t}+\left( \mathbf{u}\cdot \mathbf{\nabla }\right) \mathbf{E}=4\pi
\mathbf{j}_{eff}=4\pi (\mathbf{j+u(\rho -\mathbf{u\cdot j})}).
\label{ELLIS} \ea which, following the approach of Ref.
\cite{QFMODEL} and assuming that in momentum space
$\mathbf{u}=f(\omega)\mathbf{k}$, can be written in the form
 (\ref{GENMAXW1}-\ref{GENMAXW2}) via constitutive relations which read%
\begin{eqnarray}
\mathbf{H} &=&\mathbf{B}-f(\omega)\mathbf{k}\times \mathbf{E},  \qquad  \mathbf{D}
=\left( 1-f^{2}(\omega)k^2\right) \mathbf{E}+f^{2}(\omega)\mathbf{k}\left(
\mathbf{k}\cdot \mathbf{E}\right) -f(\omega)\mathbf{k}\times \mathbf{B}.
\label{CRELLIS}
\end{eqnarray}%
Taking $\mathbf{u}$ as a vector, this model conserves parity and
shows no birefringence.

\subsection{Myers-Pospelov Electrodynamics}

This case corresponds to the equations (\ref{MP}). From the last
one we can infer the constitutive relations \be
\mathbf{H}=\mathbf{B}-{\tilde \xi}
\partial _{0}\mathbf{E}, \qquad \mathbf{D}=\mathbf{E-}{\tilde \xi} \partial
_{0}\mathbf{B}, \label{CRMP}
\ee
which produce
\begin{equation}
\nabla \cdot \mathbf{E=}\nabla \cdot \mathbf{D,}
\end{equation}%
leaving the third Eq. (\ref{MP}) in desired form. Similarly to the
GP case, this model violates parity. In momentum space Eqs.
 (\ref{CRMP}) become \be \mathbf{H}=\mathbf{B}+i{\tilde \xi}\omega
\mathbf{E}, \qquad \mathbf{D}=\mathbf{E}+i{\tilde \xi}\omega
\mathbf{B}, \label{MPCR} \ee \vskip .3cm The above constitutive
relations in the three representative models involve linear
relations among
the fields and can be summarized in momentum space as local relations%
\begin{eqnarray}
D_{i}(\omega ,\mathbf{k})=\alpha _{ij}(\omega ,\mathbf{k})E_{i}(\omega ,%
\mathbf{k})+\rho _{ij}(\omega ,\mathbf{k})B_{j}(\omega ,\mathbf{k}),&&\nonumber  \\
H_{i}(\omega ,\mathbf{k})=\beta _{ij}(\omega ,\mathbf{k})B_{i}(\omega ,%
\mathbf{k})+\sigma _{ij}(\omega ,\mathbf{k})E_{j}(\omega
,\mathbf{k}), \label{CRCOMP}
\end{eqnarray}%
where the corresponding momentum dependent coefficients can be
read from the equations (\ref{GPCE}),\ (\ref{CRELLIS}), and
(\ref{MPCR}). Equations (\ref{CRCOMP}) are the most general
expressions in which any pair of linear constitutive relations can
be ultimately written, because one must be able to solve the
fields $\mathbf{D,H}$ in terms of $\mathbf{E,B}$.

\section{Parameterization of the constitutive relations}

Let us consider Maxwell equations in momentum space%
\begin{eqnarray}
&& \mathbf{k}\cdot \mathbf{B}\left( \omega
,\mathbf{k}\right)=0,\qquad\qquad\quad
\mathbf{k}\times \mathbf{E}\left( \omega ,\mathbf{k}\right) =\omega \mathbf{B%
}\left( \omega ,\mathbf{k}\right), \label{HOM} \\
&&i\mathbf{k}\cdot \mathbf{D}\left( \omega ,\mathbf{k}\right)=4\pi
\rho \left( \omega ,\mathbf{k}\right),\quad i\mathbf{k}\times
\mathbf{H}\left(
\omega ,\mathbf{k}\right) =-i\omega \mathbf{D}\left( \omega ,\mathbf{k}%
\right) \;+4\pi \mathbf{j}\left( \omega ,\mathbf{k}\right).
\label{INHOM}
\end{eqnarray}
Here we discuss the vacuum situation where the non trivial
constitutive relations arise because \ of LIV effects. Let us take
into account corrections up to second order in the LIV parameter
${\tilde \xi}$ and let us assume that we are in a Lorentz frame
where we demand invariance under rotations. This would correspond
to the rest frame $V^{\mu }=(1,\mathbf{0})$ in the MP model, for
example. We can always go to an arbitrary frame by means of an
observer Lorentz
transformation. In this way we have the general expressions%
\begin{eqnarray}
\alpha _{ij}=\alpha _{0}\delta _{ij}+i\alpha _{1}{\tilde \xi}
\epsilon _{irj}k_{r}+\alpha _{2}{\tilde \xi} ^{2}k_{i}k_{j},
\qquad \rho _{ij}= \rho _{0}\delta _{ij}+i\rho_{1}{\tilde \xi}
\epsilon _{irj}k_{r}+\rho _{2}{\tilde \xi}
^{2}k_{i}k_{j}, &&   \\
\beta _{ij} =\beta _{0}\delta _{ij}+i\beta _{1}{\tilde \xi}
\epsilon _{irj}k_{r}+\beta _{2}{\tilde \xi} ^{2}k_{i}k_{j}, \qquad
\sigma_{ij}=\sigma_{0}\delta _{ij}+i\sigma _{1}{\tilde \xi}
\epsilon _{irj}k_{r} +\sigma _{2}{\tilde \xi}^{2}k_{i}k_{j},&&
\end{eqnarray}%
where $\alpha _{A},\beta _{A},\rho _{A},\sigma _{A}, A=0,1,2$, are
escalar functions depending only upon $\omega$, $k=|\mathbf{k}|$,
and ${\tilde \xi}$. The property $\mathbf{k\cdot B}=0$ sets $\beta
_{2}=\rho _{2}=0$ effectively. In vector notation we then have
\begin{eqnarray}
&& \mathbf{D}=\left(\alpha_0+\alpha _{2}k^2{\tilde \xi}
^{2}\right)\mathbf{E} +\left(\rho_0+i\alpha _{1}\omega {\tilde
\xi}\right)\mathbf{B}+
i\left(\rho _{1}-i\alpha _{2}\omega{\tilde \xi}\right){\tilde \xi} \;\mathbf{k\times B},\label{D} \\
&& \mathbf{H}=\left(\sigma_0+\sigma _{2}k^2{\tilde \xi}
^{2}\right)\mathbf{E} +\left(\beta_0+i\sigma _{1}\omega {\tilde
\xi}\right)\mathbf{B} +i\left(\beta _{1}-i\sigma _{2}\omega{\tilde
\xi}\right){\tilde \xi} \;\mathbf{k\times B},\label{H}
\end{eqnarray}
where we have used the second Eq. (\ref{HOM}) together with $
\left( \mathbf{k\cdot E}\right) \mathbf{k}=\omega \left(
\mathbf{k}\times \mathbf{B}\right) \;\mathbf{+\;}k^{2}\mathbf{E}
$. Next we substitute in Eqs. (\ref{INHOM})  to obtain the corresponding equations for $\mathbf{E\;}$and $%
\mathbf{B}$. The result is
\begin{eqnarray}
&&i\left( \alpha _{0}+\alpha _{2}k^{2}{\tilde \xi} ^{2}\right) \left( \mathbf{k}\cdot \mathbf{E}%
\right) =4\pi \rho,  \label{INHOM1}\\
&&i\left( \alpha_{0}+\alpha_{2}k^{2}{\tilde \xi}^{2}\right)\omega
\mathbf{E}+i\left[ \beta_{0}+i\left( \sigma _{1}+\rho _{1}\right)
\omega {\tilde \xi} +\alpha _{2}{\tilde \xi} ^{2}\omega
^{2}\right] \mathbf{k}\times \mathbf{B}
\nonumber \\
&&\qquad\qquad\qquad\qquad+i\left[
\left(\sigma_{0}+\rho_0\right)\omega+i\left(
\alpha _{1}\omega^2-\beta_{1}k^{2}\right){\tilde \xi}\right]%
\mathbf{B}= 4\pi \mathbf{j}\left( \omega ,\mathbf{k}\right).
\label{INHOM2}
\end{eqnarray}%
Let us rewrite the inhomogeneous equations in the compact form
\begin{eqnarray}
iP\left( \mathbf{k}\cdot \mathbf{E}\right)& =&4\pi \rho,
\label{CINHOM1}\\
i\omega P\, \mathbf{E}+iQ\,\mathbf{k}\times \mathbf{B}+
R\,\mathbf{B}&=&4\pi \mathbf{J}\left( \omega ,\mathbf{k}\right),
\label{CINHOM2}
\end{eqnarray}%
by defining
\begin{equation}
P=\alpha _{0} +\alpha _{2}{\tilde \xi} ^{2}k^{2},\; Q=
\beta_{0}+i\left( \sigma _{1}+\rho _{1}\right) \omega {\tilde \xi}
+\alpha _{2}{\tilde \xi} ^{2}\omega ^{2} ,\; R=\left( \beta _{1}
k^{2}-\alpha _{1}\omega^2 \right) {\tilde \xi}+i\left(\sigma_{0}
+\rho_0\right)\omega. \label{PQR}
\end{equation}%
Now we have only three independent functions which depend on
$\omega$ and $k$.

\section{Potentials and fields}

Starting from (\ref{HOM}), (\ref{CINHOM1}) and (\ref{CINHOM2}) we
introduce the standard potencials $\Phi ,\;\mathbf{A}$
\be
\mathbf{B}=i\mathbf{k}\times \mathbf{A}, \qquad \mathbf{E}=i\omega
\mathbf{A-}i\mathbf{k} \Phi.
\ee
Substituting in the inhomogeneous equations we have%
\begin{eqnarray}
 -\omega \mathbf{k}\cdot \mathbf{A}+k^{2}\Phi &=&4\pi
\rho/P,\label{EQPHI0}\\
\left( k^{2}Q-\omega ^{2}P\right) \mathbf{A}+i R\;\left( \mathbf{k}%
\times \mathbf{A}\right) - \left[ Q\left( \mathbf{k}\cdot
\mathbf{A}\right)-P\omega \;\Phi \right] \mathbf{k}&=&4\pi
\mathbf{j}. \label{EQA0}
\end{eqnarray}
We can easily fix two convenient gauges as follows.

\subsection{Generalized Lorentz gauge}

In the standard situation the Lorentz gauge in momentum space is \
$\mathbf{k}\cdot \mathbf{A}=\omega \Phi$.  Here we choose
\begin{equation}
 \mathbf{k}\cdot \mathbf{A} =
\left(P/Q\right) \omega \Phi ,  \label{GLG}
\end{equation}%
in such a way that Eqs. (\ref{EQA0}) and (\ref{EQPHI0}) reduce to
\begin{eqnarray}
&&\left( Qk^{2}-P\omega ^{2}\right) \mathbf{A}+i R\;\left( \mathbf{k}%
\times \mathbf{A}\right) = 4\pi \mathbf{j}  \label{AGLG}\\
&&\left( Qk^{2}-P\omega ^{2}\right) \Phi =4\pi\left(Q/P\right)
\rho, \label{PHIGLG}
\end{eqnarray}
respectively.

\subsection{The radiation gauge}

As usual we set $ \mathbf{k}\cdot \mathbf{A=}0 $
and we obtain the equations%
\begin{eqnarray}
\Phi &=&4\pi\left( k^{2}P\right)^{-1} \rho,  \label{PHIRG}\\
\left( Qk^{2}-P\omega ^{2}\right) \mathbf{A}+i R\;\left( \mathbf{k}%
\times \mathbf{A}\right) &=&4\pi \mathbf{j}-4\pi \omega \rho
k^{-2} \mathbf{k}.  \label{ARG}
\end{eqnarray}%
Eq. (\ref{PHIRG}) exhibits the scalar potential as a static
contribution, which subsequently does not contribute to the
radiation field. The use of current conservation in (\ref{ARG})
allows the introduction of the transversal current, consistently
with the chosen gauge. That is to say we have
\begin{equation}
\left( Qk^{2}-P\omega ^{2}\right) \mathbf{A}+i kR\;\left( \mathbf{%
\hat{k}}\times \mathbf{A}\right) =4\pi \left[ \mathbf{j-}\left(
\mathbf{j}\cdot \mathbf{\hat{k}}\right) \mathbf{\hat{k}}\right]
=4\pi \mathbf{j}_{T}.\label{EQARG}
\end{equation}%
The presence of birefringence depends on the term proportional to
$kR$, where we have a crucial factor of $i$. This makes clear
that a possible diagonalization can be obtained by using a
complex basis, which is precisely the circular polarization
basis. In fact,
decomposing the vector potential and the current in such basis%
\begin{equation}
\mathbf{A}=\mathbf{A}^{+}+\mathbf{A}^{-},\;\;\;\;\mathbf{j}_{T}=\mathbf{j}%
_{T}^{+}+\mathbf{j}_{T}^{-},
\end{equation}%
and recalling the basic properties%
\begin{equation}
\mathbf{\hat{k}}\times \mathbf{A}^{+}=-i\mathbf{A}^{+},\;\;\mathbf{\hat{k}}%
\times \mathbf{A}^{-}=i\mathbf{A}^{-},
\end{equation}%
we can separate (\ref{EQARG}) into the uncoupled equations
\begin{equation}
\left[ Qk^{2}-P\omega ^{2}+\lambda  k\,R%
\right] \mathbf{A}^\lambda =4\pi \mathbf{j}_{T}^{\lambda}
,\;\;\;\;\lambda =\pm 1.
\end{equation}

\section{Green functions in the radiation gauge}

To our aims it is convenient to work in this gauge, where the
polarized Green function is defined by
\begin{equation}
\left( Q k^{2}-P \omega ^{2}+\lambda k\,R \;\right) G^{\lambda
}(\omega ,k) =1.
\end{equation}%
For each polarization, the above leads to a cubic equation in $k$,
which determines the dispersion relations for the propagating
photon. In fact one of the poles is due to the $\beta_1$ factor in
the expression for $R$ in Eq. (\ref{PQR}). This pole is at $k\sim
\xi^{-1}$, and therefore its contribution can be neglected in our
effective theory valid for $k\ll \xi^{-1}$. We defer for future
work a detailed characterization of the solutions and proceed here
on the basis of general properties of light propagation, which
constitute a generalization of the specific models already
considered. We characterize the propagation mode corresponding to
each polarization $\lambda$ by a refraction index
$n_\lambda(\omega)$ to be read from the appropriate dispersion
relation in such a way that
$n_\lambda(\omega)={k}_\lambda(\omega)/\omega$. The general form
of the polarized Green function is
\begin{equation}
G^{\lambda}(\omega,{\mathbf r})=\frac{1}{4\pi r}F^{\lambda}
(\omega) e^{i\omega n_{\lambda}\left(  \omega\right)r}\label{nbr},%
\end{equation}
where $\hat\mathbf n=\mathbf{r}/r$ is the direction of
observation. The generic form for the  total  Green function in
the far-field approximation is
\begin{equation}
G_{ret}^{ij}\left(  \omega,{\mathbf r} \right)=\frac{1}{2}\left[\left(  \delta^{ik}\mathbf{-}\hat{n}%
^{i}\hat{n}^{k}+i\epsilon^{irk}\hat{n}^{r}\right)  G^{+}\left(
\omega ,{\mathbf r}\right)  +\left(
\delta^{ik}\mathbf{-}\hat{n}^{i}\hat{n}^{k}-i\epsilon
^{irk}\hat{n}^{r}\right)  G^{-}\left(\omega,{\mathbf r} \right)\right]. \label{br}%
\end{equation}
Note that from the birefringent case we can go to the
non-birefringent one by taking $
n_{+}(\omega)=n_{-}(\omega)=n(\omega)$, in which case  $
F^{+}(\omega)=F^{-}(\omega)=F(\omega)$
and the Green function (\ref{br}) becomes%
\begin{equation}
G_{ret}^{ik}\left(  \omega,{\mathbf r} \right)  =\frac{1}{4\pi
r}\left( \delta ^{ik}\mathbf{-}\hat{n}^{i}\hat{n}^{k}\right)
F\left(\omega \right)
e^{i\omega n(\omega) r}.%
\end{equation}
In the following we analyze the main consequences of an
electrodynamics characterized by a Green function of the type
(\ref{br}). To warrant that the
fields are real, it must be%
\begin{equation}
[G^{+}\left(  \omega,{\mathbf r}\right)]^{\ast}=G^{-}\left(
-\omega,{\mathbf r}\right).
\end{equation}
{ This implies the relations}
\begin{equation}
n_{+}^{\ast}(\omega)=n_{-}(-\omega), \qquad
[F^{+}(\omega)]^{\ast}=F^{-}(-\omega). \label{eo}
\end{equation}
For a birefringent medium the real and imaginary parts of the
refraction index for circular polarization components can contain
both $\omega$-even and $\omega$-odd terms, provided that they
satisfy Eq. (\ref{eo}). In the case of a non-birefringent medium
the real part of the refraction index is even in $\omega$, and the
imaginary part is odd. We can see that the refraction indices for
the Myers-Pospelov theory, Eq. (\ref{REFIND0}), satisfy these
requirements.

The Green functions must also be consistent with causality, which
means that they cannot have poles in the $\omega$ upper half
plane. This leads to generalized Kramers-Kronig
relations, which for a dispersive refraction index imply the
existence of an imaginary part. This means that these media  are
necessarily absortive. If we assume that these theories are valid
in a range of frequencies where the effective medium can be
considered as transparent, then in the case of  birefringent
theories we have a real refraction index with both $\omega$-even
and $\omega$-odd terms, while for non-birefringent theories we
have only $\omega$-even terms.

Next we turn to the calculation of the angular distribution of the
power spectrum. The full vector potential is given by the
superposition
\begin{equation}
\mathbf{A}(\omega,\mathbf{r})=\frac{1}{r}\sum_{\lambda=\pm1}F^{\lambda
}(\omega)e^{in_{\lambda}\omega r}\mathbf{j}^{\lambda}(\omega,\mathbf{{k}_{\lambda}%
})=\mathbf{A}_{+}+\mathbf{A}_{-}.\label{VECPOT}%
\end{equation}
where $\mathbf{k}_{\lambda}=n_\lambda(\omega)\ \omega
\mathbf{\hat{n}}$ and  $\mathbf{\hat{n}}=\mathbf{r}/r$. From here
we can compute the
electric and magnetic fields%
\begin{eqnarray}
\mathbf{B(}\omega,\mathbf{r})  & =&\mathbf{\nabla}\times\mathbf{A(}%
\omega,\mathbf{r})=\frac{\omega}{r}\sum_{\lambda=\pm1}\lambda n_{\lambda
}F^{\lambda}e^{in_{\lambda}\omega r}\mathbf{j}^{\lambda}(\omega,\mathbf{{k}%
_{\lambda}})=\omega\left(
n_{+}\mathbf{A}_{+}-n_{-}\mathbf{A}_{-}\right),
\label{EFIELD}\\
\mathbf{E(}\omega,\mathbf{r})  & =&i\omega\mathbf{A(}\omega,r)=\frac{i\omega
}{r}\sum_{\lambda=\pm1}F^{\lambda}e^{in_{\lambda}\omega r}\mathbf{j}^{\lambda
}(\omega,\mathbf{{k}_{\lambda}})=i\omega\left(  \mathbf{A}_{+}+\mathbf{A}%
_{-}\right).
\end{eqnarray}
The Poynting vector
\begin{eqnarray}
\mathbf{S}(\omega, {\mathbf r})
&=&\frac{1}{4\pi}\mathbf{E}(-\omega,
{\mathbf r})\times\mathbf{H}(\omega, {\mathbf r}),  \nonumber \\
&=&\frac{\omega^2}{4\pi}\sum_{\lambda=\pm1} \left\{i\left[
\left(\beta_0+i\sigma_1\omega{\tilde \xi}\right)
n_\lambda+i\lambda\sigma_0\right] \left[ \mathbf{A}_+\left(
\lambda\omega,\mathbf{r}\right) \times \mathbf{A}_-\left(-\lambda
\omega,\mathbf{r}\right)\right] \right. \nonumber \\
&&\qquad\qquad\left.+\lambda\beta_1{\tilde \xi}\omega n_\lambda^2
\left[ \mathbf{A}_+\left( \lambda\omega,\mathbf{r}\right) \cdot
\mathbf{A}_-\left(-\lambda
\omega,\mathbf{r}\right)\right]\hat\mathbf{n}\right\},
\end{eqnarray}
allows us to compute the angular distribution of the energy spectrum
\begin{eqnarray}
\frac{d^{2}E}{d\Omega d\omega}  &
=&\frac{r^{2}}{2\pi}\mathbf{\hat{n} }\cdot\left[
\mathbf{S}\left(  \omega,r\right)  +\mathbf{S}%
\left(  -\omega,r\right)  \right],\nonumber  \\
& =&\frac{r^{2}\omega^{2}}{8\pi^{2}} \left[
Z(\omega)\,\mathbf{A}_{-}\left( -\omega,\mathbf{r}\right)
\cdot\mathbf{A}_{+}\left( \omega,\mathbf{r}\right)
+Z(-\omega)\,\mathbf{A}_{-}\left( \omega,\mathbf{r}\right)
\cdot\mathbf{A}_{+}\left( -\omega,\mathbf{r}\right)
\right],\label{Ee}
\end{eqnarray}
where
\begin{equation}
Z(\omega)=\sum_{\lambda=\pm1}\left\{i\sigma_0\left(\lambda\omega\right)
+\left[\frac{}{} \beta_0(\lambda\omega)+i\lambda\,
\sigma_1(\lambda\omega)\, \omega{\tilde \xi}\right]
n_\lambda(\lambda\omega)+\beta_1(\lambda\omega)\, {\tilde \xi}
\omega\,  n^2_\lambda(\lambda\omega)\right\}. \label{Z}
\end{equation}
Using Eqs. (\ref{eo}) and (\ref{VECPOT}), together with the
general property $ [\mathbf{j}^\lambda(\omega,
\mathbf{k})]^{\ast}=\mathbf{j}^{-\lambda}(-\omega,
-\mathbf{k}^{\ast})$, which allows for  complex refraction
indexes,
we finally get%
\begin{equation}
\frac{d^{2}E}{d\Omega d\omega }=\frac{\omega ^{2}}{8\pi ^{2}}%
\sum_{\lambda =\pm }Z(\lambda \omega )\left| F^{\lambda }(\omega
)\right| ^{2}e^{-2\omega \rm{Im}\left[ n_{\lambda }(\omega \right]
r}\;\left[ \mathbf{j}^{\lambda }(\omega ,\mathbf{k}_{\lambda
}(\omega ))\right] ^{\ast }\cdot \mathbf{j}^{\lambda }(\omega
,\mathbf{k}_{\lambda }(\omega )),\label{EEAD}
\end{equation}
where the vectors $\mathbf{k}_{\lambda }(\omega )$ were defined
after Eq.(\ref{VECPOT}). As expected, absorption is present
through the non-zero imaginary part of the refraction indexes.
Eq.(\ref{EEAD}) is related with the angular distribution of the
radiated power spectrum by
\begin{equation}
\frac{d^{2}E}{d\Omega d\omega }=\int
dT\;\frac{d^{2}P(T)}{dwd\Omega }, \label{ep}
\end{equation}
where $T$ is a macroscopic time, which leads to \ba
\frac{d^{2}P(T)}{d\Omega d\omega}&=&\frac{\omega ^{2}}{8\pi ^{2}}%
\sum_{\lambda =\pm }\left[Z(\lambda \omega )\left| F^{\lambda
}(\omega )\right| ^{2}e^{-2\omega {\rm Im}\left[ n_{\lambda
}(\omega \right]
r}\right.\; \nonumber \\
&& \left. \frac{}{}\times \int_{-\infty}^{\infty}d\tau\
e^{-i\omega\tau} {j}_k^{\ast }(T+\tau/2 ,\mathbf{k}_{\lambda
}(\omega ))\,\, P^\lambda_{kr}\,\, {j}_r(T+\tau/2
,\mathbf{k}_{\lambda }(\omega ))\right], \ea where $
P^\lambda_{kr}=1/2\left(\frac{}{} \delta_{kr}- {\hat n}_k\,{\hat
n}_r + i \lambda \epsilon_{klr}{\hat n}_l \right)$ are the
circular polarization projectors.

In the sequel we consider synchrotron radiation in the situation
where the vacuum absorption can be neglected so that the
refraction indices are real and Eq. (\ref{eo}) reduces to
$n^+(\omega)=n^-(-\omega)$. Also we assume that
$F^\lambda(\omega)$ and $Z^\lambda(\omega)$ are real functions.
Using the current corresponding to the circular
orbits in (\ref{EQCHARGE}) we can compute the cycle average over the macroscopic time $T$ obtaining%
\begin{equation}
\left\langle \frac{d^{2}P(T)}{d\omega d\Omega}\right\rangle =
\sum_{\lambda=\pm1}\sum_{m=0}^{\infty}\delta({\omega}-m\omega_{0}%
)\frac{dP_{m,\lambda}}{d\Omega},%
\end{equation}%
with
\begin{equation}
\frac{dP_{m\lambda }}{d\Omega }=\frac{\omega _{m}^{2}q^{2}}{8\pi }%
Z(\lambda \omega )\left[ \frac{F^{\lambda }(\omega )}{n_{\lambda }(\omega )}%
\right] ^{2}\left[ \lambda \beta n_{\lambda }(\omega
)J_{m}^{\prime }(W_{m\lambda })+cot\theta \,J_{m}(W_{m\lambda
})\right] ^{2},
\end{equation}
where $W_{m\lambda}=m n^\lambda(\omega)\beta\sin \theta$. The total averaged power
radiated in the $m^{th}$ harmonic%
\begin{equation}
P_{m}=\frac{q^{2}\beta ^{2}\omega _{m}}{2R}\sum_{\lambda =\pm
1}\left\{
Z(\lambda \omega )\frac{[F^{\lambda }(\omega )]^{2}}{n^{\lambda }(\omega )}%
\left[ J_{2m}^{\prime }(2m\beta n_{\lambda })-\frac{1-\beta
^{2}n_{\lambda }^{2}}{2 \beta ^{2}n_{\lambda
}^{2}}\int_{0}^{2m\beta n_{\lambda }}dx\;J_{2m}(x)\right]
\right\},
\end{equation}
where the contribution of each polarization is exhibited.

In the case of the Myers and Pospelov effective theory the
non-null coefficients in the constitutive relations
(\ref{D}-\ref{H}) are
\begin{equation}
\alpha_0=\beta_0=1,\qquad \sigma_0=\rho_0=i{\tilde \xi}\omega,
\end{equation}
while the remaining functions are given by \be
F^\lambda(\omega)=\frac{n_\lambda(\omega)}{\sqrt{1+({\tilde
\xi}\omega)^2}}, \qquad Z(\omega)=2\sqrt{1+({\tilde
\xi}\omega)^2}. \ee From the general expressions obtained in this
section we recover the results of Section 2.

\section{Final remarks}
We have presented a compact review of the most interesting results
for synchrotron radiation in the Myers-Pospelov LIV
electrodynamics. This is one of several existing models of LIV
electrodynamics that also include those proposed by
Gambini-Pullin, and Ellis et. al., for example. Motivated by such
diversity we have developed a general approach based on
generalized constitutive relations, set out in terms of the
standard methods for describing  radiation in a dispersive and
absortive medium with a given index of refraction. The resulting
formalism has been explicitly applied to synchrotron radiation in
the case of transparent media.

\ack LFU  acknowledges the organizers of the School for their
invitation to participate and  also the partial support from the
projects CONACYT-M\'exico-40745-F and DGAPA-UNAM-IN104503-3. RM
acknowledges partial support from CONICET-Argentina.

\end{document}